\documentclass[letterpaper]{article} % DO NOT CHANGE THIS
\usepackage{aaai2026}  % DO NOT CHANGE THIS
\usepackage{times}  % DO NOT CHANGE THIS
\usepackage{helvet}  % DO NOT CHANGE THIS
\usepackage{courier}  % DO NOT CHANGE THIS
\usepackage[hyphens]{url}  % DO NOT CHANGE THIS
\usepackage{graphicx} % DO NOT CHANGE THIS
\usepackage{natbib}  % DO NOT CHANGE THIS AND DO NOT ADD ANY OPTIONS TO IT
\usepackage{caption} % DO NOT CHANGE THIS AND DO NOT ADD ANY OPTIONS TO IT
\usepackage{algorithm}
\usepackage{algorithmic}

\usepackage{booktabs}  
\usepackage{array}   
\usepackage{multirow}
\usepackage{amsmath}
\usepackage[table]{xcolor}

\title{One Rule to Bring Them All: Investigating Transport Connectivity in Public Transport Route Generation for Equitable Access}

\author {
    Aleksandr Morozov\textsuperscript{\rm 1},
    Ruslan Kozliak\textsuperscript{\rm 1},
    Georgii Kontsevik\textsuperscript{\rm 1},
    Sergey Mityagin\textsuperscript{\rm 1}
}

\affiliations {
    \textsuperscript{\rm 1}ITMO University, St. Petersburg, Russia\\
    mityagin@itmo.ru 
}

\begin{document}

\maketitle

\begin{abstract}
Designing a city-wide public transport network poses a dual challenge: achieving computational efficiency while ensuring spatial equity for different population groups. We investigate whether AI-based optimization — hybrid neuroevolutionary methods combining graph neural networks with evolutionary algorithms — can scale Transit Network Design Problem (TNDP) solutions from synthetic tests to real urban networks while preserving social fairness. Our contribution is to introduce a transport connectivity-aware accessibility metric that bases optimization on principles of equitable accessibility rather than traditional trade-offs between passenger and operator costs. The results show a noticeable improvement in network resilience by improving algebraic connectivity on synthetic datasets, and highlight the ambiguity of applying network generation to real data.

\end{abstract}

\section{Introduction}
To build more equitable and sustainable cities, we need to recognize the new demands posed by new urban science. Both new methods of studying and modeling the urban environment and the emergence of new forms of cities open up new perspectives for studying existing cities and assessing the consequences of new urban phenomena \cite{ye2025artificial}. Artificial intelligence and new technologies are causing rapid changes in urban planning, creating new functional urban spaces and changing the purpose of existing ones. Changes in the structure of locations and the recombination of functions are occurring under the spatial influence of innovations in ICT \cite{jacques2024smart}. Such digital innovations in the field of transport, for example, are leading to the emergence of driverless transport/autonomous vehicles (AVs), which is leading to the repurposing of public parking spaces due to the reduction in the use of private cars \cite{batty2024digital}. However, as spatial interventions follow the application of new technologies, the algorithms that tell us how to optimally design cities need to be studied in more detail to ensure spatial equity for different groups of residents \cite{soydan2025user}. In the field of transportation, this may mean citizens' right to equitable access to any important facilities, such as hospitals, business centers, or kindergartens \cite{hook2025evaluating}.

% Much work has been done in the field of researching reliable and trustworthy AI, particularly to mitigate bias in AI decision-making. However, few studies acknowledge these implications in an urban context, especially with regard to equitable transportation and access. The rise of autonomous vehicles, shared mobility solutions, and advances in public transportation are occurring alongside rapid changes in urban structure. Recent computational advances in this area include the use of deep learning-based solutions, where the design of reward functions is of paramount importance. Thus, as AI-based solutions have a growing direct impact on our lives in the physical space of cities, it is crucial to design such systems in a way that delivers socially equitable and sustainable solutions. 

Planning a public transport system involves several optimization criteria, which makes it difficult to select a globally optimal structure \cite{borndorfer2017line}. An effective system must meet the functional requirements of both passengers and operators, while ensuring accessibility and comfort for all residents \cite {borndorfer2017line}. Automating the planning process without using additional parameters (such as transport frequency schedules or taking into account the synergy of several modes of transport in a single network) can be considered as a TNDP (transport network design problem). This task is computationally complex due to the extensive search space and list of domain constraints \cite{mumford2013new}. Given the complexity of the problem, researchers have applied various heuristic methods and agent-based modeling to find the optimal network configuration \cite{holliday2024nea}.

Being in the field of mathematical problems, many works exploring TNDP solutions have been applied to synthetic graphs, with a few practical examples for different cities.

However, the transition from synthetic to real networks reveals serious problems that go beyond accelerating computations. From an urban planning perspective, a number of methodological questions arise regarding mathematically optimal solutions \cite{da2025generativeaitransportationplanning}. While AI reduces the time needed to find solutions for real cities \cite{holliday2024nea}, synthetic tests (Mandl, Mumford) do not take into account the spatial heterogeneity of POI distribution and demographic complexity characteristic of real urban conditions \cite{rodriguez2025analysis} to ensure a socially acceptable result.

%  Performance metrics related to customer satisfaction (customer costs), such as the number of transfers required (studied less frequently) or average travel time (studied more frequently), do not fully reflect spatial equity. 

Such solutions typically do not provide sufficient service to nodes with lower demand, while the balance between operator and customer costs can lead to noticeable shortcomings in customer service. Existing studies have shown that an algorithm priority aimed at complete customer satisfaction using the average travel time indicator lead to a huge increase in operating costs, making such solutions unfeasible in real-world conditions.

This paper answers the question of whether attempting to provide equitable access to facilities for a a bigger number of residents can be sustainable in terms of operator and customer costs, and whether there are any advantages to such an approach.

The contribution of this work is the introduction of an accessibility metric as a target optimization function based on the principle of fair accessibility. This approach has three significant advantages. First, it provides a near-optimal solution, avoiding the search for the optimal balance of customer and operator weights. Second, the generated network demonstrates better resilience to disruptions. Third, although continuous accessibility itself remains close to the original solution, a more reliable transport network promises better spatial equality.

\section{Related Work}

The planning process in the field of public transport is conceptually divided into several stages, the results of which serve as input for the subsequent stage. \citet{schmidt2024planning} decompose this process into three major subproblems:

\begin{itemize}
    \item Line generation,
    \item Line selection,
    \item Frequency setting.
\end{itemize}

In this work, we focus on the stages of line generation and selection, formally known as the line planning or Transit Network Design Problem (TNDP) \citep{mandl1980,duran2022survey, csahin2023line}. The objective of this stage is to determine a feasible set of transit lines that ensures convenient travel for passengers while minimizing operational costs. The infrastructure network is represented as a graph, where nodes correspond to stops and edges denote the physical or temporal connections between them, weighted by distance or travel time. Each transit line is represented as an ordered sequence of stops. The inputs to this problem typically include the underlying infrastructure network and a travel demand matrix, which is generally assumed to be static over the analysis period \citep{schmidt2024planning}. Extensions of this problem often include temporal or multimodal dimensions, such as frequency optimization \citep{jha2019multi} or re-routing strategies aimed at improving gravity-based accessibility without increasing operational costs \citep{rumpf2021public}.

A substantial body of research on TNDP relies on metaheuristic methods, which are well suited to handling its combinatorial nature and flexible objective formulations. Among these, evolutionary algorithms have gained particular prominence, including genetic algorithms designed with specialized operators for TNDP \citep{bagloee2011transit, pternea2015sustainable}, and their multi-objective variants, such as NSGA-II \citep{liu2020pareto}, which balance passenger convenience and operator costs.

Recent research has increasingly explored the use of machine learning and neural network architectures capable of capturing the topological dependencies within transport graphs. In particular, graph neural networks (GNNs) and reinforcement learning (RL) methods enable modeling the process of route construction as a sequence of actions in a Markov decision environment. Several studies have demonstrated the potential of deep neural architectures in combinatorial optimization tasks structurally similar to TNDP, such as the Traveling Salesman Problem (TSP) and Vehicle Routing Problem (VRP), where GNN models are trained to generate graph-based solutions that account for spatial interdependencies \citep{khalil2017learning, wu2021learning}. Building upon these advances, \citet{holliday2024nea} proposed integrating a GNN model trained in the RL paradigm with an evolutionary algorithm for the construction and improvement of public transport networks, showing that such hybrid approaches can effectively combine structural reasoning with adaptive optimization.

Despite notable progress in algorithmic methods for solving the TNDP, most studies remain primarily focused on improving computational efficiency and enhancing network-level or passenger performance metrics. Comparatively few works have addressed the question of how generated transit networks influence spatial accessibility and equity across urban territories.

\citet{wang2024public} highlight the importance of incorporating spatial accessibility and geographic equity metrics into network optimization frameworks. Their model, based on Message Passing Neural Networks (MPNN) and reinforcement learning, aims to minimize disparities in access to urban opportunities, demonstrating that aligning network topology with the spatial and functional structure of the city can support the design of more equitable transport systems. However, such approaches remain limited, and the comprehensive evaluation of generated networks in terms of equal access to spatial units within the city is still an underexplored research direction.

\section{Methodology}

Our methodological framework is designed to compare different route generation strategies under a unified formulation of the Transit Network Design Problem. All candidate solutions are represented as feasible sets of routes on the same underlying city graph, subject to identical route length and connectivity constraints. 

Each solution is evaluated using a composite objective that combines three components: passenger cost ($C_p$), operator cost ($C_o$), and a demand-weighted transport connectivity metric ($C_w$), complemented by penalty terms for infeasibility. By varying the weights assigned to these components, we emulate different planning perspectives (user-oriented, operator-oriented, and connectivity-oriented) while preserving a consistent evaluation scheme.

Within this common setting, we benchmark three algorithms, developed by \citet{holliday2024nea}: a learned constructive baseline (LC100), a metaheuristic baseline (BCO), and a neuro-evolutionary variant (NeuroBCO) that integrates a learned policy into the search process. All methods are run on the same benchmark instances with identical parameters, enabling a controlled assessment of how the choice of objective and optimization strategy affects network efficiency, connectivity, and the equity of access.

\subsection{Problem Formulation}

The Transit Network Design Problem is formulated on an augmented city graph
\[
C = (N, E_s, D),
\]
where $N$ denotes the set of $n$ potential transit stops (nodes), $E_s = \{(i,j,\tau_{ij})\}$ is the set of street edges with associated travel times $\tau_{ij} > 0$, and $D \in R_{\ge 0}^{n \times n}$ is a symmetric demand matrix ($D = D^\top$) specifying the number of trips between each origin–destination pair~\cite{mandl1980}. 

A \emph{route} $r$ is a sequence of distinct nodes connected by edges in $E_s$. The objective is to determine a set of routes $\mathcal{R} = \{r_1, \dots, r_S\}$ that collectively form a feasible transit network, satisfying
\[
\text{MIN} \le |r| \le \text{MAX},
\]
for all $r \in \mathcal{R}$, such that all nodes are mutually reachable via transit, and a multi-objective cost function is minimized. This cost typically balances passenger travel efficiency, operational expenditure, and network feasibility constraints. The TNDP is NP-hard due to the combinatorial nature of route selection and the interdependence of demand coverage and transfer penalties.

In this paper, we carry out experiments with generating sets of routes for several benchmark graphs with different configurations. The input consists of graphs where the vertices are public transport stops and the edges are paths between stops with weights in times (minutes). LC-100, EA, and NEA algorithms are used for generation (description of the algorithms is provided below). For each experiment, the following optimization functions were selected: Average Travel Time, Route Travel Time, and Weighted Transport Connectivity.

\subsection{Neuro-evolutionary algorithm}

The Neural-Evolutionary Algorithm (NEA)~\cite{holliday2024nea} is a hybrid optimization method for the TNDP that integrates a reinforcement learning-based graph neural network (GNN) policy with an evolutionary algorithm (EA). The learned policy acts as a heuristic mutation operator, allowing the algorithm to explore the solution space more efficiently than purely random search strategies.

At the core of NEA lies a GNN policy $\pi_\theta(a|s)$, referred to as the \emph{Learned Constructor}. It constructs a complete set of transit routes by interacting with an environment represented as a Markov Decision Process (MDP), where the city graph $C=(N, E_s, D)$ defines the state space. At each step, the policy decides whether to extend or terminate a route based on node and edge embeddings produced by a graph attention network. The cumulative reward corresponds to the negative of the overall cost function:
\begin{equation}
C(C,\mathcal{R}) = \alpha w_p C_p + (1-\alpha)w_o C_o + \beta C_c,
\end{equation}
where $C_p$ is the passenger cost, $C_o$ the operator cost, and $C_c$ a constraint penalty term. The parameters $\alpha$, $w_p$, $w_o$, and $\beta$ balance the influence of each component.

The passenger cost represents the average in-vehicle travel time across all demand pairs:
\begin{equation}
C_p(C, \mathcal{R}) = 
\frac{\sum_{i,j} D_{ij} \tau_{ij}^{\mathcal{R}}}
{\sum_{i,j} D_{ij}},
\end{equation}
where $\tau_{ij}^{\mathcal{R}}$ denotes the shortest travel time between $i$ and $j$ along the transit network $\mathcal{R}$, including a fixed penalty $p_T$ per transfer.  

The operator cost quantifies total route traversal time:
\begin{equation}
C_o(C, \mathcal{R}) =
\sum_{r \in \mathcal{R}} \tau_r, 
\quad \tau_r = \sum_{(u,v) \in r} \tau_{uv} + \sum_{(v,u) \in r} \tau_{vu}.
\end{equation}

The constraint-violation penalty $C_c$ is defined as
\begin{equation}
    C_c = f_u + 0.1[f_u > 0] + f_l + 0.1[f_l > 0],
\end{equation}
where $f_u$ is the fraction of positive-demand OD pairs with no path in the planned route set $R$, 
$f_l$ is the total route-length violation (how far all routes deviate above MAX or below MIN stops), normalized to the scale of a minimum-length route, 
MIN and MAX are the prescribed lower and upper bounds on route length (in number of stops), 
and $[P] = 1$ if the condition $P$ is true and $0$ otherwise.
% The penalty term $C_c$ accounts for disconnected node pairs and violations of route length constraints.

During evolutionary optimization, a population of candidate networks undergoes iterative mutation and selection. In addition to standard random mutations, NEA employs a \emph{neural mutation}, where the GNN policy $\pi_\theta$ generates a new route to replace an existing one in the current solution. This hybridization combines the exploratory power of evolutionary search with the problem-specific knowledge embedded in the learned policy. The resulting framework achieves improved convergence and solution quality compared to classical EAs and standalone learned models, performing effectively on standard TNDP benchmarks.

\subsection{Transport connectivity}

The transport connectivity metric is based on the principle of mutual accessibility of urban areas and reflects how effectively the transport network enables movement between all pairs of spatial units (e.g., urban blocks). It is computed by determining the minimum travel time between all nodes in the intermodal transport network, normalized by the corresponding straight-line (Euclidean) distances \citep{morozov2023assessing}. 

To account for the spatial distribution of demand, a demand-weighted component is introduced. Let $T^\mathcal{R} \in R^{n \times n}$ denote the matrix of all-pairs shortest transit times. The weighted median connectivity is then defined as:
\begin{equation}
C_w(C, \mathcal{R}) = \frac{1}{n} \sum_{i=1}^{n} 
\operatorname{median}\left\{ T_{ij}^\mathcal{R} \cdot 
\frac{D_{ij}}{\max_k D_{ik}} \;\middle|\; T_{ij}^\mathcal{R} < \infty \right\}.
\end{equation}

This formulation emphasizes accessibility to high-demand destinations and complements traditional cost-based optimization criteria by linking the structure of the transport network with the spatial distribution of travel needs, allowing for a more comprehensive evaluation of urban connectivity.

\section{Experiments and Results}
\subsection {Data}

All \textbf{experiments}\footnote{\url{https://anonymous.4open.science/r/TNDP_learning-EC76/experiments.ipynb}} were conducted on the standard public transit benchmarks \citep{mumford_utrp}: the Mandl network (15 nodes, 6 routes) and the four larger Mumford instances (ranging from 30 to 127 nodes and 12 to 60 routes), as detailed in Table~\ref{tab:synthetic_cities}. These synthetic cities vary significantly in scale, demand distribution, and spatial structure, providing a robust testbed for evaluating transit network generation algorithms under diverse conditions. The algorithms under comparison - LC-100, BCO, and the hybrid NeuroBCO - were executed with a fixed random seed and identical route length constraints to ensure reproducibility and fair performance assessment.

\begin{table}[htbp]
\centering
\caption{Statistics of the five synthetic benchmark cities used in our experiments.}
\label{tab:synthetic_cities}
\begin{tabular}{lrrrrrr}
\toprule
\textbf{City} & \textbf{$n$} & \textbf{$|E_s|$} & \textbf{$S$} & \text{min/max} & \text{Area (km$^2$)} \\
\midrule
Mandl     & 15  & 20  & 6  & 2 / 8   & 352.7 \\
Mumford0  & 30  & 90  & 12 & 2 / 15  & 354.2 \\
Mumford1  & 70  & 210 & 15 & 10 / 30 & 858.5 \\
Mumford2  & 110 & 385 & 56 & 10 / 22 & 1394.3 \\
Mumford3  & 127 & 425 & 60 & 12 / 25 & 1703.2 \\
\bottomrule
\end{tabular}
\end{table}

\subsection{Assessment}

In the access-graph methodology of \citet{vsfiligoj2025access}, accessibility evaluation begins with constructing a generalized travel-time matrix \(D\), defined as

\begin{equation}
D_{ij} =
T^{\mathrm{veh}}_{ij}
+ w_{\mathrm{wait}} \cdot T^{\mathrm{wait}}_{ij}
+ w_{\mathrm{trans}} \cdot N_{\mathrm{trans}}(i,j),
\label{eq:gen_travel_time}
\end{equation}

Two network representations are used to obtain its components. The L-space encodes physical adjacency between successive stops and supplies the in-vehicle travel times used to compute the shortest-path travel time \(T^{\mathrm{veh}}_{ij}\). The P-space links all stops served by the same route, enabling the extraction of route sequences and identification of transfer counts \(N_{\mathrm{trans}}(i,j)\). In the original formulation, expected waiting times are frequency-based, the waiting-time perception weight is \(w_{\mathrm{wait}}=2\), and each transfer contributes a 5-minute penalty.

Because benchmark networks in this study do not provide service frequencies, the waiting-time component is approximated by a fixed 5-minute value per transfer and the perception weight is set to \(w_{\mathrm{wait}}=1.0\). Transfers are determined directly from changes in route identifiers in P-space paths. These adjustments preserve the behavioral logic of generalized travel impedance while ensuring compatibility with the available data.

For each travel-time budget \(tb\), an access graph \(G_A(tb)\) is constructed by forming an adjacency matrix \(A_{ij}(tb)\) whose elements indicate whether two stops can reach each other within the given generalized travel-time threshold:

\begin{equation}
A_{ij}(tb) =
\begin{cases}
1, & \text{if } D_{ij} \le tb, \\[2pt]
0, & \text{otherwise}.
\end{cases}
\label{eq:adjacency_matrix}
\end{equation}

As \(tb\) increases, more stop pairs satisfy the threshold, producing progressively denser graphs. The time budget is systematically varied from very small values up to the maximum generalized travel time \(t_{\max}\) using a 2-minute increment. Special cases include \(tb=30\) minutes, representing a standard benchmark for typical one-way urban travel, and the characteristic budget \(t_M\), corresponding to the point of most rapid connectivity expansion.

For each access graph \(G_A(tb)\), the average degree is computed as

\begin{equation}
D(tb) = \frac{1}{n} \sum_{i=1}^{n} \deg_{G_A(tb)}(i),
\label{eq:avg_degree}
\end{equation}

where \(\deg_{G_A(tb)}(i)\) is the degree of node \(i\) in the adjacency-defined access graph. The sequence \(D(tb)\) is non-decreasing and describes how the number of reachable destinations grows as additional travel time becomes permissible. To identify the characteristic budget \(t_M\), the discrete first difference of the average degree is evaluated:

\begin{equation}
\Delta D(tb_k) = D(tb_k) - D(tb_{k-1}),
\label{eq:first_difference}
\end{equation}

and divided by the budget increment to obtain the discrete growth rate \(\Delta D/\Delta tb\). The value \(t_M\) is then defined as the \(tb_k\) at which this growth rate reaches its global maximum, marking the steepest increase in accessibility and indicating the transition from localized clusters to widespread network connectivity.

Based on the evolution of adjacency-defined access graphs \(G_A(tb)\), the following indicators are extracted:
\begin{itemize}
    \item \textbf{\(t_M\)} — the travel-time budget with the steepest increase of the average degree, representing the onset of rapid accessibility expansion (see \eqref{eq:avg_degree} and \eqref{eq:first_difference}).  
    \item \textbf{\(\delta t_M = t_M / t_{\max}\)} — normalized characteristic time, indicating how quickly the network becomes substantially connected relative to its scale.  
    \item \textbf{\(D_M\)} — average degree at \(t_M\) (see \eqref{eq:avg_degree}), quantifying the number of reachable opportunities at the point of fastest connectivity growth.  
    \item \textbf{\(D_{30}\)} — average degree at 30 minutes (see \eqref{eq:avg_degree}), providing a standard accessibility measure for typical urban travel.  
    \item \textbf{\(G_M\)} — Gini coefficient of degrees at \(t_M\), measuring spatial equity of accessibility during the critical transition.  
    \item \textbf{\(G_{30}\)} — Gini coefficient at 30 minutes, evaluating the equity of medium-range accessibility.
\end{itemize}

Together, the generalized travel-time model \eqref{eq:gen_travel_time}, the adjacency-defined access graphs \eqref{eq:adjacency_matrix}, and the systematic sweep over travel-time budgets produce a comprehensive framework for characterizing the dynamics of connectivity formation, the scale of reachable destinations at meaningful thresholds, and the equity of accessibility across the network.

\subsection{Experiments on Benchmark Data}

The core of our analysis revolves around the demand-weighted connectivity metric $C_w$, which integrates spatial accessibility, travel efficiency, and demand responsiveness into a single objective. Unlike traditional metrics that focus solely on average travel time ($C_p$) or total route length ($C_o$), $C_w$ captures how effectively the network connects urban areas in proportion to real travel needs. Results across all benchmarks (Tables 1--5 in additional materials, consolidated in Table 6 in additional materials) reveal that $C_w$ acts as a natural mediator between passenger and operator objectives, preventing extreme trade-offs and promoting more balanced, equitable transit systems.

When optimization prioritizes passenger convenience (high $C_p$ weight), networks achieve the lowest average travel times by constructing dense, overlapping routes around high-demand corridors. While this minimizes in-vehicle time and transfer penalties, it comes at a steep operational cost: $C_o$ often increases by 2-3 times compared to cost-focused solutions. For example, on the Mandl benchmark, passenger-optimized configurations reduce $C_p$ to approximately 10.4 minutes but inflate $C_o$ beyond 350 minutes of total route traversal. This pattern scales with network size-on Mumford3, $C_o$ exceeds 8700 under passenger-focused settings-highlighting the economic unsustainability of purely user-centric designs.

In contrast, minimizing operator cost generates compact, tree-like structures with minimal route mileage. These networks are highly efficient from an operational standpoint, achieving the lowest $C_o$ across all instances, but at the expense of passenger experience. Direct-trip shares ($d_0$) plummet - often below 15\% in larger networks - and multi-transfer journeys dominate, with over 30\% of trips requiring two or more changes in Mumford3. Such configurations create accessibility bottlenecks, particularly in low-demand peripheral areas, resulting in spatially inequitable service.

The introduction of $C_w$ as a primary or balanced objective fundamentally alters this dynamic. Networks optimized with significant $C_w$ weighting exhibit substantially improved transfer profiles: direct trips rise sharply (up to 70\% in balanced settings on smaller graphs), and long transfer chains are nearly eliminated (see Figure 3 in additional materials). On Mumford0, for instance, $d_0$ increases from just 15.5\% under operator minimization to nearly 70\% in connectivity-aware configurations, with two-transfer trips dropping to under 1\%. This enhancement in direct connectivity occurs without fully sacrificing efficiency - $C_o$ remains within 20-50\% of the operator-optimal baseline in most cases.

Balanced weight combinations (e.g., equal or moderate emphasis on $C_p$, $C_o$, and $C_w$) consistently produce solutions near the Pareto front, achieving near-minimal $C_w$ while preserving competitive values in both $C_p$ and $C_o$. These configurations demonstrate the complementary nature of the three objectives: connectivity-driven routing naturally induces redundancy that benefits passengers (shorter paths, fewer transfers) while avoiding the excessive sprawl of pure passenger optimization. Figure 1 and 2 in additional materials illustrates this synergy, showing normalized metric trade-offs across weight settings for NeuroBCO on Mumford3.

Among the tested algorithms, NeuroBCO demonstrates superior performance when $C_w$ is included in the objective. It consistently achieves the lowest $C_w$ values - often by margins of 5--8\% on larger instances - while maintaining strong $C_p$ scores. This advantage stems from its hybrid architecture, which combines learned graph-aware construction with evolutionary refinement, enabling more nuanced exploration of demand-structured connectivity patterns than purely heuristic or evolutionary baselines.

Despite these benefits, over-emphasizing $C_w$ reveals important limitations. Exclusive connectivity optimization increases $C_o$ by 20--40\% relative to balanced settings, as the algorithm introduces redundant or overlapping routes to maximize weighted accessibility. In datasets with uneven demand (common in Mumford instances), this can lead to overfitting: high-demand clusters become over-served, while peripheral nodes risk disconnection or poor service. Moreover, a ``connectivity paradox'' emerges---beyond a moderate threshold, further reductions in $C_w$ no longer improve $C_p$ and may even degrade it due to inefficient path structures and increased route density.

These findings underscore that while $C_w$ is a powerful tool for promoting equitable, demand-responsive transit networks, its optimal impact is realized in moderation. Balanced configurations - particularly those assigning non-zero weights to all three objectives - consistently deliver robust, accessible, and operationally viable solutions. They avoid the pitfalls of single-objective extremes and produce networks that are not only efficient and user-friendly but also resilient and spatially just.
\subsubsection{Comparative Analysis on Pareto Front and Accessibility Metrics}

\begin{figure}[htbp]
    \centering
    \includegraphics[width=0.95\linewidth]{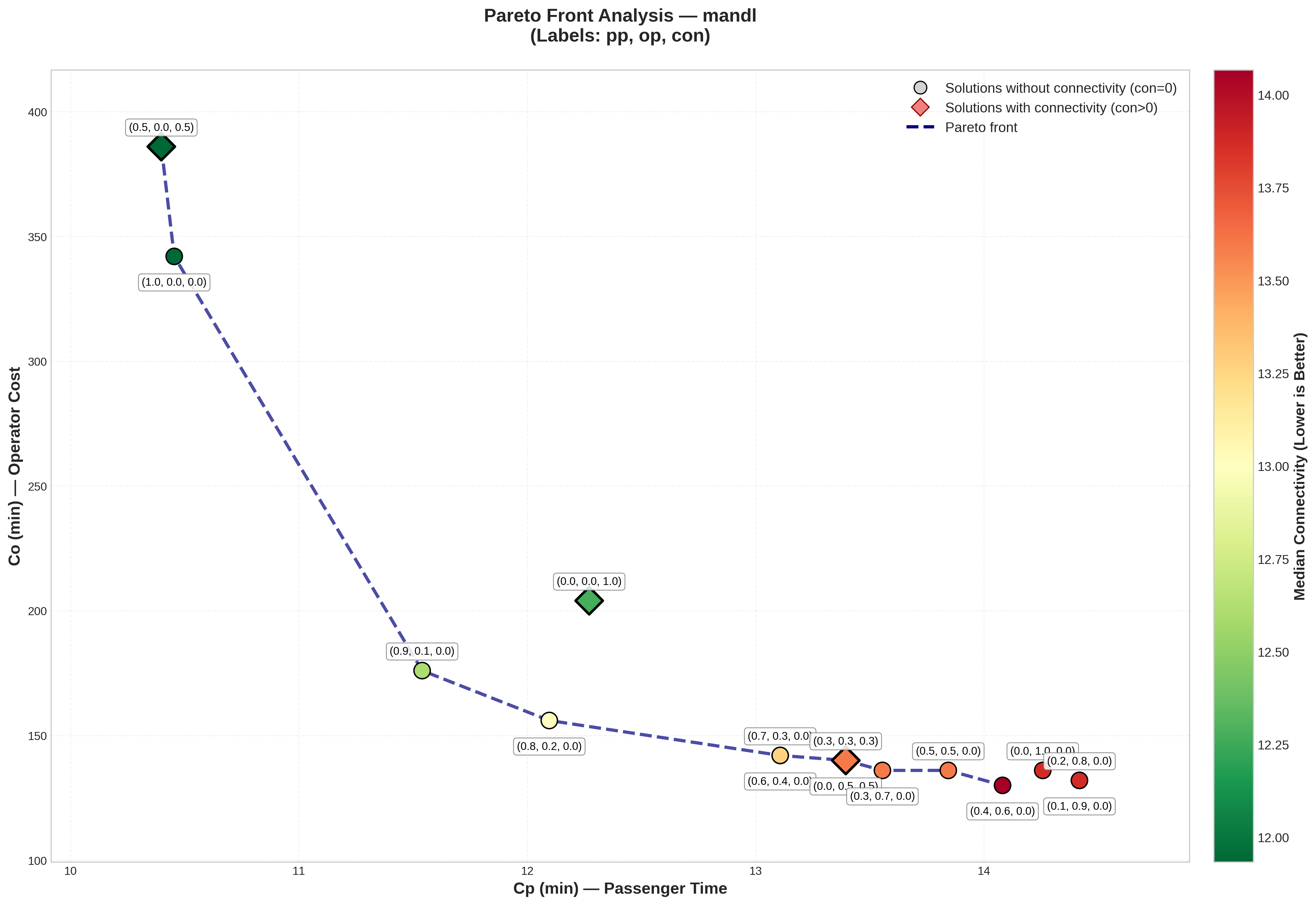}\\[1em]
    \includegraphics[width=0.45\linewidth]{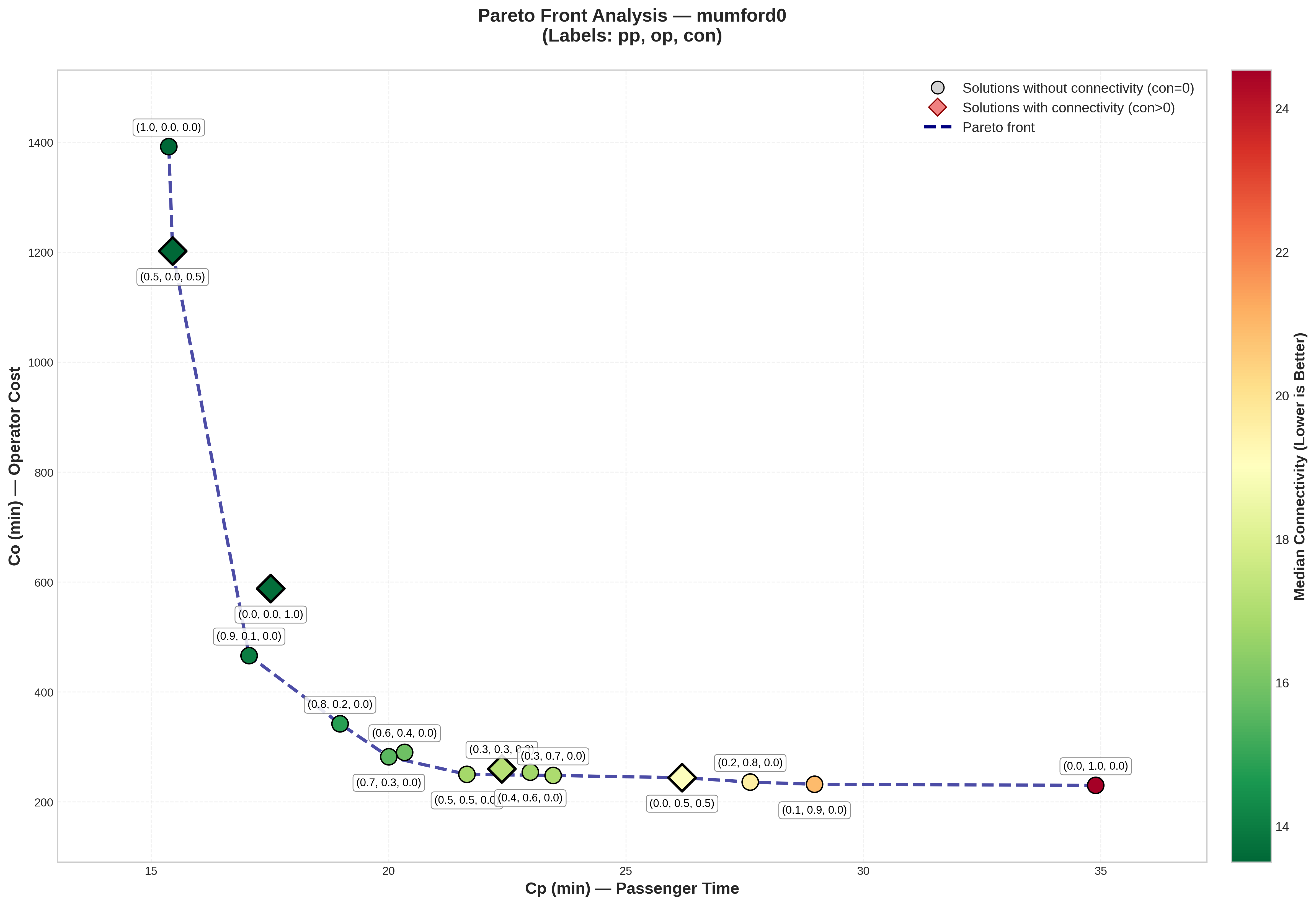}
    \includegraphics[width=0.45\linewidth]{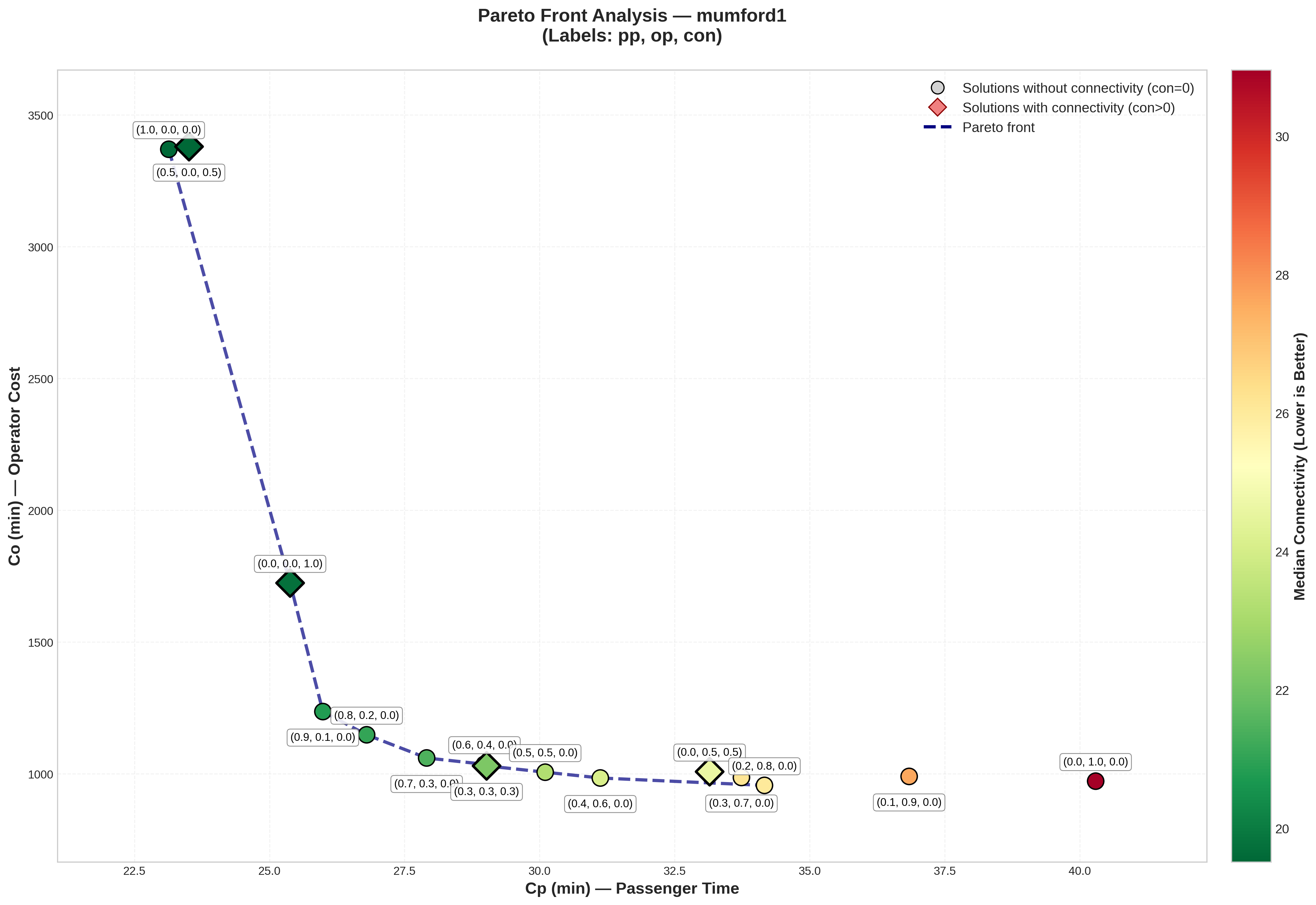}\\[1em]
    \includegraphics[width=0.45\linewidth]{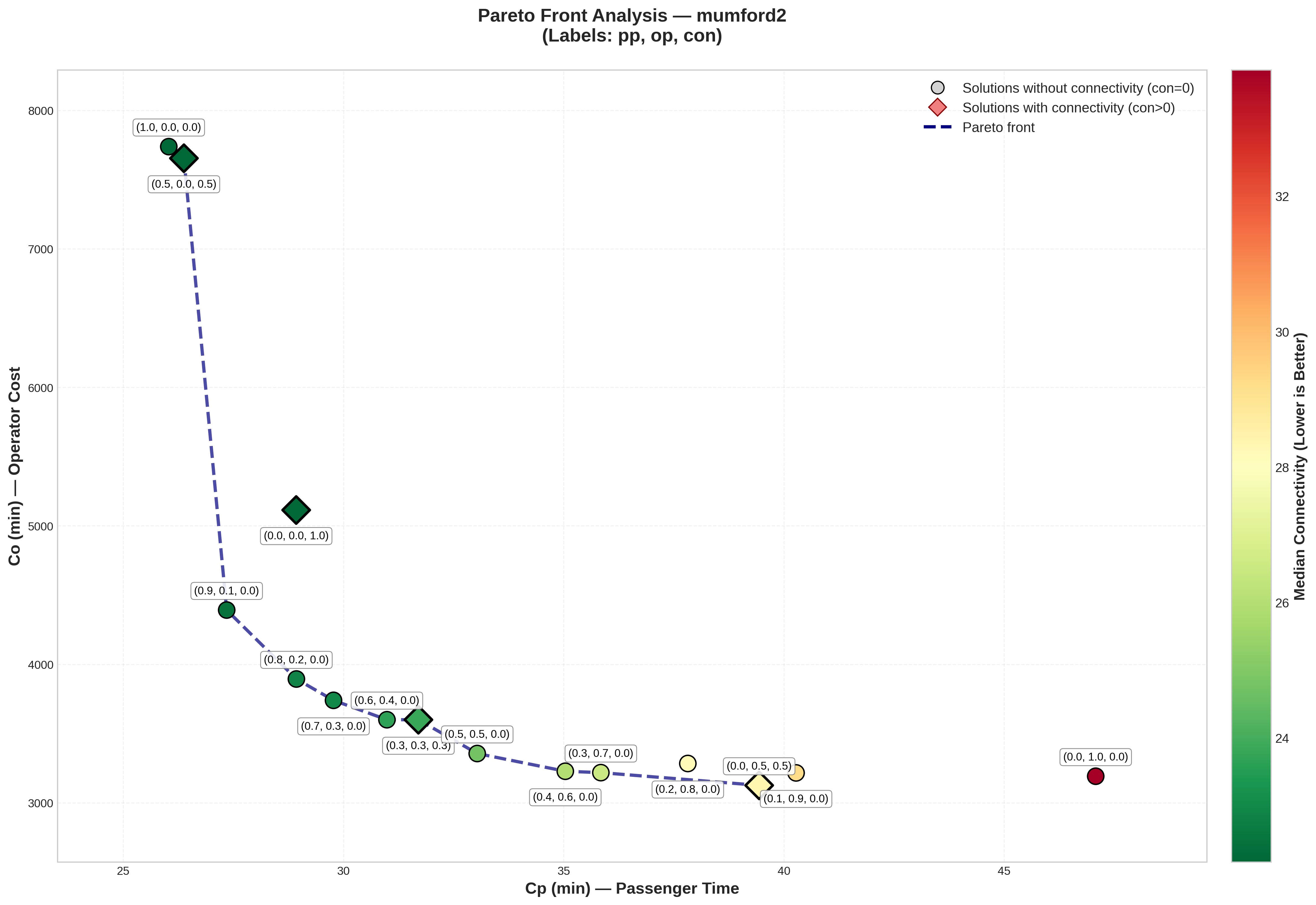}
    \includegraphics[width=0.45\linewidth]{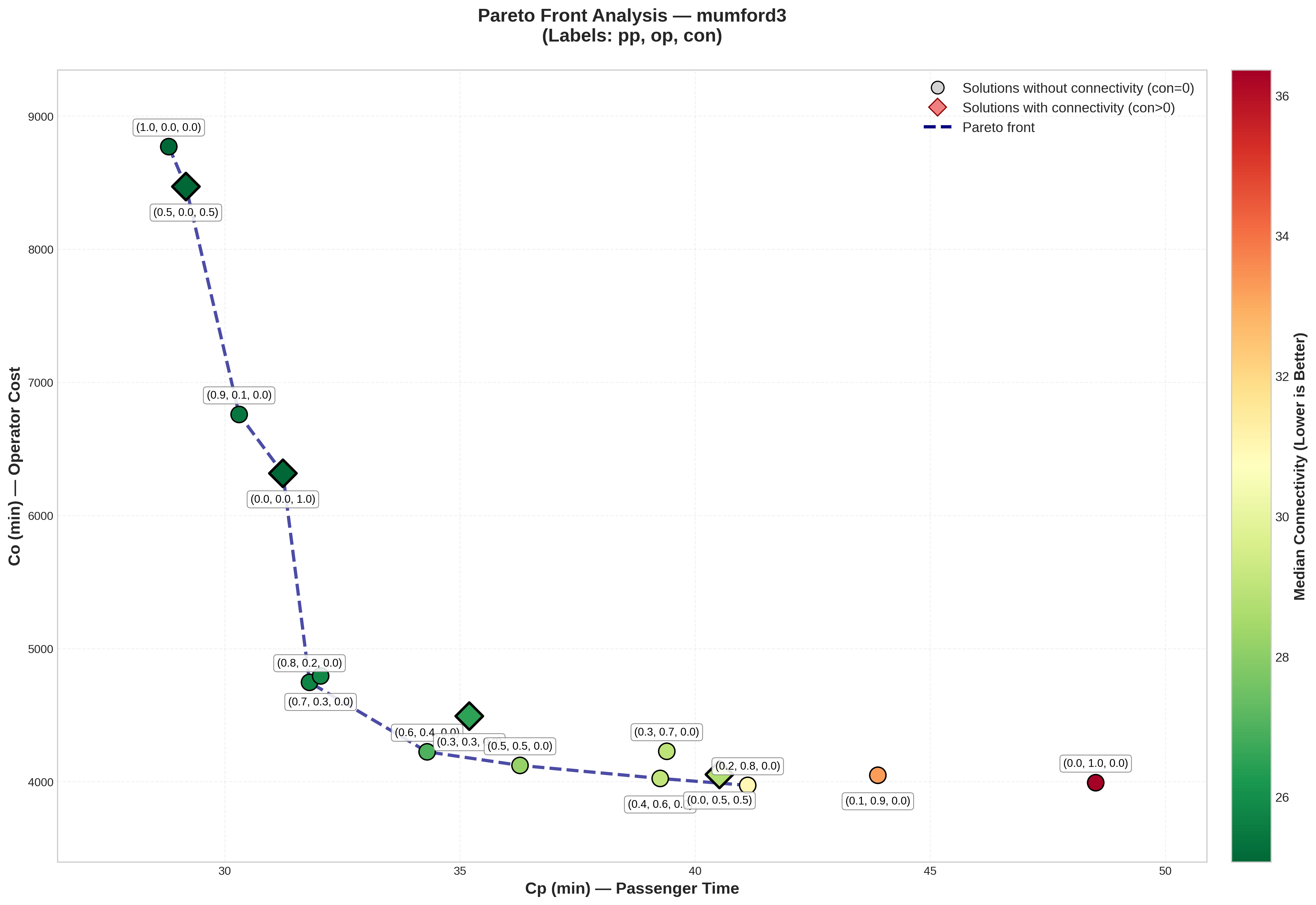}
    \caption{Pareto fronts of passenger cost $C_p$ and operator cost $C_o$ across benchmark instances Mandl and Mumford0--3. The rhomb marks our solution for $w>0$.}
    \label{fig:pareto_fronts_combined}
\end{figure}

To deepen the understanding of the trade-offs in our transit network optimization, we analyze the solutions on the Pareto front illustrated in Figure~\ref{fig:pareto_fronts_combined}. The Pareto front combines results from diverse coefficient settings, highlighting our solution at point (0,0,1). Notably, the solution at coefficient set (0,0,1) is closely situated to (0.9,0.1,0), warranting a focused comparison alongside the extreme points at (0,1,0) and (1,0,0). These points effectively represent the emphasis on connectivity, operator cost, and passenger convenience respectively.

\begin{table*}[tb]
  \centering
  \small
  \caption{Merged Results: performance metrics for coefficient settings on Mumford2 benchmark.
           \textit{AC} refers to Algebraic Connectivity, a spectral measure of network
           robustness and connectivity, reflecting the equity and resilience of transit
           access across the city. The algorithm was run with fixed seed and route length constraints as in Table~\ref{tab:synthetic_cities}. All transfer-related metrics ($d_{>2}$, $d_0$, $d_1$, $d_2$) are reported in percentages, where $d_{>2}$ denotes the share of trips requiring more than two transfers.}
  \label{tab:merged_results}
  \begin{tabular*}{\textwidth}{@{\extracolsep{\fill}} lccccccccccccccc @{}}
    \toprule
    \textbf{Weights}
    & $t_{max}$↓ & $t_{M}$↓ & $\delta t_{M}$↓ & $D_{M}$↑ & $D_{30}$↑ & $G_{M}$↓ & $G_{30}$↓
    & AC↑ & $C_p$↓ & $C_o$↓ & $C_{con}$↓
    & $d_0$ & $d_1$ & $d_2$ & $d_{>2}$ \\
    \midrule
    1.0 0.0 0.0
      & 76.0 & 24.0 & 0.32 & 47.6 & 69.4 & 0.15 & 0.12
      & \cellcolor{green!25}1.01
      & \cellcolor{green!25}26.0
      & \cellcolor{red!25}7738.0
      & \cellcolor{green!25}22.2
      & 42.6 & 55.6 & 1.8 & 0.0 \\
    0.9 0.1 0.0
      & 74.0 & 24.0 & 0.32 & 46.0 & 67.3 & 0.16 & 0.13
      & \cellcolor{orange!25}0.77
      & \cellcolor{yellow!25}27.4
      & \cellcolor{yellow!25}4392.0
      & \cellcolor{orange!25}22.5
      & 28.3 & 62.5 & 9.1 & 0.1 \\
    0.0 1.0 0.0
      & 264.0 & 30.0 & 0.11 & 26.9 & 26.9 & 0.24 & 0.24
      & \cellcolor{red!25}0.08
      & \cellcolor{red!25}47.1
      & \cellcolor{green!25}3192.0
      & \cellcolor{red!25}33.9
      & 11.4 & 24.1 & 28.0 & 36.5 \\
    \rowcolor{gray!15}
    0.0 0.0 1.0
      & 71.0 & 28.0 & 0.39 & 57.4 & 64.1 & 0.14 & 0.14
      & \cellcolor{yellow!25}0.96
      & \cellcolor{orange!25}28.9
      & \cellcolor{orange!25}5114.0
      & \cellcolor{green!25}22.2
      & 21.5 & 52.2 & 24.6 & 1.7 \\
    \bottomrule
  \end{tabular*}
\end{table*}

Our solution, corresponding to the weight configuration (0,0,1), is highlighted on the Pareto fronts in Figure~\ref{fig:pareto_fronts_combined}. Although positioned close to the (0.9,0.1,0) configuration — which slightly emphasizes passenger convenience with minimal operator-cost consideration—the connectivity-focused (0,0,1) solution shows clear advantages across the accessibility indicators proposed by \citet{vsfiligoj2025access}.

Most notably, our solution achieves a higher algebraic connectivity (AC = 0.96 vs. 0.77), reflecting a more robust and resilient network structure. This improvement in spectral connectivity supports more balanced service distribution and greater resistance to disruptions, contributing directly to equitable access across different parts of the city.

This benefit is further reflected in the access-graph-derived metrics. The characteristic time $t_M$ at which the network rapidly becomes accessible is slightly higher than in passenger-leaning settings, but the relative characteristic time $\delta t_M$ remains moderate, indicating efficient scaling of connectivity. Reachability at the critical growth point is stronger than in the (0.9,0.1,0) case, meaning more nodes become accessible during the key transition to full network connectivity. Equity, as measured by the Gini coefficients of node degree distributions, is also better—showing a more even spread of access opportunities and reducing disparities between high-demand and peripheral areas.

The maximum travel time is lower than in passenger-optimized configurations and significantly better than operator-only minima, further limiting exposure to long or inequitable journeys. Together, these results demonstrate that prioritizing weighted connectivity not only improves network robustness but also enhances performance across the dynamic and equity-sensitive indicators introduced by \citet{vsfiligoj2025access}. Even when solutions appear similar on the traditional $C_p$–$C_o$ Pareto front, our connectivity-driven approach generates networks that are structurally stronger, more balanced, and better aligned with the goal of equitable urban accessibility.

\subsubsection{Routes visualization}
\begin{figure}[htbp]
    \centering
    \includegraphics[width=1\linewidth]{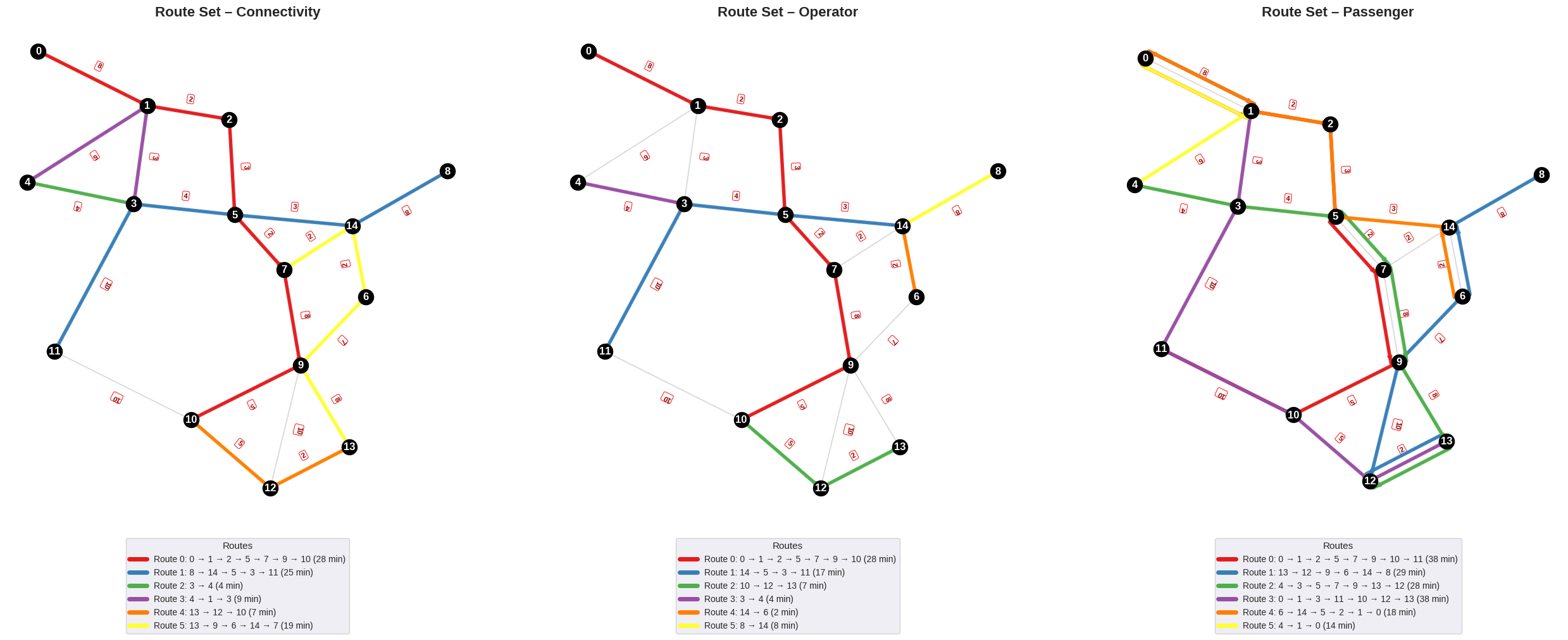}
    \caption{Transit route sets in the Mandl benchmark optimized under three objectives: (left) connectivity, (center) operator cost, and (right) passenger convenience. The operator-perspective solution (center) uniquely forms a \textbf{spanning tree} of the full Mandl graph, containing exactly 14 edges, full vertex coverage, and guaranteed connectivity without redundancy.}
    \label{fig:routes}
\end{figure}

Figure~\ref{fig:routes} illustrates the transit route sets in the Mandl benchmark optimized under three perspectives: connectivity (left), operator cost (center), and passenger convenience (right).

The connectivity-focused route set (left) prioritizes maximum coverage and network robustness, ensuring that every stop is reachable from every other stop with high redundancy. It consists of 6 routes and forms a highly interconnected graph with multiple overlapping paths between key nodes. This structure guarantees full vertex coverage and strong connectivity, even in case of disruptions, but includes numerous cycles and redundant edges, reflecting a design that sacrifices efficiency for resilience and accessibility.

The resulting set of routes (center) forms a spanning tree of the full Mandl transport graph with 15 vertices. All vertices are covered, and the union of edges from the routes contains exactly 14 connections, each present in the original Mandl graph. Connectivity verification confirms that the subgraph is connected, and the absence of cycles follows from the minimal number of edges. Thus, the routes constitute a minimal connected structure that ensures reachability between any pair of stops without redundant overlaps, making the solution optimal in terms of network connectivity with the minimal number of transit lines. This optimal spanning tree structure is achieved using the coefficient combination (pp=0, op=1, cp=0), corresponding to the operator perspective, which prioritizes minimizing the number of routes while maintaining full network coverage and connectivity.

The passenger-oriented route set (right) focuses on meeting the travel demands and convenience of residents, emphasizing short travel times, frequent service, and easy transfers. It includes 6 routes with multiple parallel or overlapping connections between high-demand nodes, creating local clusters and dense central hubs. This results in redundant pathways and cycles, enabling passengers to reach nearby destinations quickly-often without transfers - but at the cost of operational efficiency. The design reflects real-world urban transit patterns, where user comfort and demand-driven routing take precedence over minimal infrastructure.

\subsection{Experiments on City Dataset}

For experiments on real network, we assembled a model of the city of Tartu (Estonia) with a population of 97,759 and an area of 38.8 km². 

\begin{figure*}[t]
    \centering
    \includegraphics[width=\linewidth]{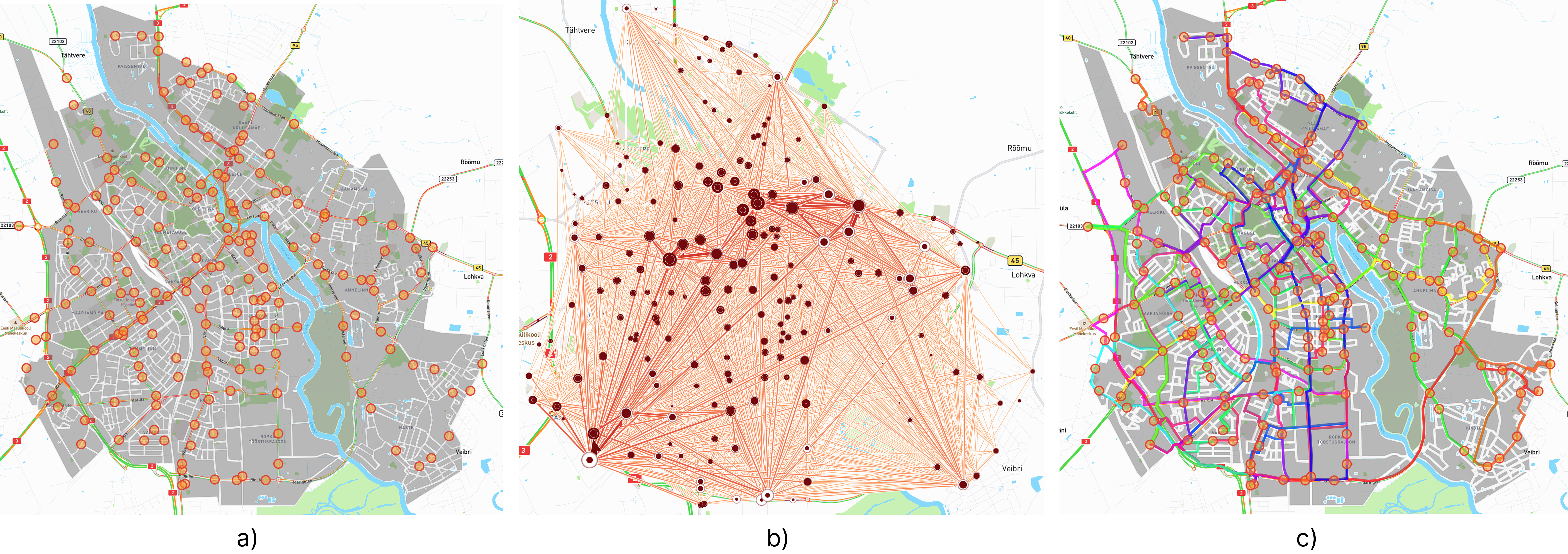}
    \caption{Visualization of results on the Tartu network: 
    (a) Preprocessed urban blocks with stops, 
    (b) OD flows (visualized using \textbf{Flowmap City}\textsuperscript{*} platform), 
    (c) Generated routes on road network for configuration  $(0,0,1)$.}
    \label{fig:tartu_results}
\end{figure*}

A proxy origin–destination (OD) matrix is constructed using a gravity model and block-level indicators that describe local population and service characteristics and gravity model (Figure \ref{fig:tartu_results}{a}). Indicators were calculated using the \textbf{BlocksNet} Python framework\footnote{\url{https://github.com/aimclub/blocksnet}}. Buildings with their parameters, roads, and city services from \textbf{OpenStreetMap (OSM)}\footnote{\url{https://www.openstreetmap.org}} were used as input data. Below is a formulation for constructing an OD matrix.

Let $Q$ denote the set of urban blocks, each characterized by population, service density, and diversity parameters $\{\text{population}_q, \text{density}_q, \text{diversity}_q\}$. The street network is represented as a graph $G=(V,E)$, where $V$ is the set of intersections and $E$ the set of road segments. Data aggregation is performed within a radius of $R=10$ minutes of walking accessibility.

For each block $q \in Q$, a land-use coefficient $\text{lu\_coeff}_q$ is assigned according to its land-use type. Based on these attributes, normalized demand and supply indicators are computed as follows:
\begin{align}
P_q &= \text{population}_q, \label{eq:Pq}\\
A_q &= \text{lu\_coeff}_q + \text{density}_q + \text{diversity}_q. \label{eq:Aq}
\end{align}

These block-level parameters are then projected onto nearby public transport stops $i \in V$ by distance-weighted aggregation:
\begin{align}
P_i &= \sum_{q : i \in S_q} \frac{P_q}{d(q,i)}, \label{eq:Pi}\\
A_i &= \sum_{q : i \in S_q} \frac{A_q}{d(q,i)}, \label{eq:Ai}
\end{align}
where $d(q,i)$ is the shortest-path distance between block $q$ and stop $i$, and $S_q$ denotes the set of stops within the accessibility radius $R$.

\begin{table*}[tb]
  \centering
  \caption{Performance metrics for coefficient settings on Tartu. 
  Transfer-related metrics ($d_{>2}$, $d_0$, $d_1$, $d_2$) are reported in percentages.}
  \label{tab:Tartu_results}
  \scalebox{0.9}{ % <-- регулируй масштаб вручную
  \begin{minipage}{1.1\textwidth} % можно немного больше 1.0
  \begin{tabular}{@{\extracolsep{\fill}} lccccccccccccccc @{}}
    \toprule
    \textbf{Weights}
    & $t_{max}$↓ & $t_{M}$↓ & $\delta t_{M}$↓ & $D_{M}$↑ & $D_{30}$↑ & $G_{M}$↓ & $G_{30}$↓
    & AC↑ & $C_p$↓ & $C_o$↓ & $C_{con}$↓
    & $d_0$ & $d_1$ & $d_2$ & $d_{>2}$ \\
    \midrule
    1.0 0.0 0.0
      & 50 & 26 & 13.87 & 73.45 & 126.92 & 0.25 & 0.17
      & \cellcolor{yellow!25}0.017
      & \cellcolor{green!25}17.9
      & \cellcolor{orange!25}1432.1
      & \cellcolor{green!25}13.2
      & 19.4 & 49.5 & 28.4 & 2.7 \\
    0.9 0.1 0.0
      & 53 & 26 & 12.84 & 67.19 & 118.46 & 0.27 & 0.19
      & \cellcolor{yellow!25}0.017
      & \cellcolor{yellow!25}18.69
      & \cellcolor{green!25}1205.4
      & \cellcolor{orange!25}13.6
      & 19.7 & 49.2 & 27.8 & 3.4 \\
    0.0 1.0 0.0
      & 62 & 30 & 10.81 & 84.96 & 84.96 & 0.22 & 0.22
      & \cellcolor{orange!25}0.009
      & \cellcolor{red!25}24.08
      & \cellcolor{yellow!25}1328.3
      & \cellcolor{red!25}16.6
      & 15.4 & 35.0 & 36.4 & 13.3 \\
    \rowcolor{gray!15}
    0.0 0.0 1.0
      & 56 & 26 & 13.25 & 69.94 & 121.58 & 0.28 & 0.19
      & \cellcolor{green!25}0.020
      & \cellcolor{orange!25}19.11
      & \cellcolor{red!25}1540.8
      & \cellcolor{yellow!25}13.5
      & 16.8 & 43.2 & 33.3 & 6.7 \\
    \bottomrule
  \end{tabular}
  \end{minipage}
  }
\end{table*}

Finally, the proxy OD matrix is constructed by combining the potential of origins and destinations, weighted by their spatial separation:
\begin{equation}
x_{ij} = \frac{P_i \, A_j}{d(i,j)}, \label{eq:xij}
\end{equation}
where $d(i,j)$ is the travel distance between stops $i$ and $j$ along the street network. The resulting matrix $X = [x_{ij}]$ serves as a synthetic approximation of travel demand, capturing the spatial interaction between population concentration and service attractiveness across the city.

The graph of the street and road network is compiled based on OSM data and preprocessed, leaving only the paths between stops. Python code from the \textbf{Transport}\footnote{\url{https://anonymous.4open.science/r/transport-3D32/examples/example_preprocess.ipynb}} repository is used for preprocessing. As weights on the edges of the graph, travel time is calculated based on a speed of 20 km/h. The public transport stops were aggregated using a model from the paper \citet{lee2012stop}.

The process of preparing data on the street and road network consists of several stages. First, a subgraph is formed containing only vertices marked as stops. Next, for each stop ($i$), Dijkstra's algorithm is performed to obtain the lengths and geometries of the shortest paths to all other stops ($j$). Only paths with exactly two stops (start and end) are considered, which excludes connections with intermediate stop nodes and forces the resulting edges to represent “clean” sections of movement without pickups/dropoffs. Additionally, a path is discarded if its total length exceeds the acceptable range or if any segment lacks a valid weight. However, it is very complicated to determine this weight, as the distance between stops may vary from city to city. Even when using the median or average value, long distance edges may be lost. For the selected pairs of stops, the weight is aggregated based on the specified speed and distance according to the edge geometry. The original list of nodes is saved as an edge attribute, which allows the internal structure of the route to be reconstructed. After processing all pairs, the graph is reduced to a compact form by renumbering the vertices. The final graph of the preprocessed network for experiments has 199 nodes and 1235 edges. OD flows by proxy matrix are shown in the (Figure \ref{fig:tartu_results}{b}).

{\renewcommand\thefootnote{}\footnotetext{\textsuperscript{*}\url{https://www.flowmap.city}}\addtocounter{footnote}{-1}}
% \footnotetext[*]{\url{https://www.flowmap.city}}

We conducted experiments on the Tartu network using a fixed configuration of 15 routes with a minimum length of 20 and a maximum length of 40 stops (nodes).
% For the experiments, a slightly different parameterization was applied. Instead of a fixed average waiting time $t_{wait}$, we defined a default service frequency of 4 trips per hour (approximately one departure every 15 minutes). The average waiting time was then derived as the half headway, computed as $t_{wait} = 30 / \textit{frequency}$, assuming uniformly distributed vehicle arrivals.
The results in Table~\ref{tab:Tartu_results} show trade-offs consistent with those observed on synthetic benchmarks, now within a real urban context.
The passenger-oriented configuration $(1,0,0)$ delivers relatively low travel times ($t_{max}=50$, $t_M=26$) and the highest reachability ($D_{30}\approx126.9$), while maintaining a favourable transfer profile with a small share of long itineraries ($d_{>2}=2.7\%$). Introducing a small operator weight in $(0.9,0.1,0)$ slightly increases travel times and $C_{con}$, but reduces operational cost ($C_o\approx1205.4$) and preserves both accessibility and transfer structure, indicating that mild cost-awareness can be incorporated without substantial loss in user performance.
In contrast, the operator-dominated configuration $(0,1,0)$ yields the weakest outcome from an accessibility perspective: travel times increase ($t_{max}=62$, $C_p\approx24.1$), algebraic connectivity drops ($AC\approx0.009$), and the share of trips with more than two transfers rises to $d_{>2}=13.3\%$, reflecting a sparser and less equitable network.
The connectivity-focused configuration $(0,0,1)$ achieves the highest algebraic connectivity ($AC\approx0.020$) and solid reachability ($D_{30}\approx121.6$), but requires the largest operational cost ($C_o\approx1540.8$) and only partially improves the transfer distribution ($d_{>2}=6.7\%$). This suggests that optimizing for connectivity alone yields structurally robust networks, but with limited practical gains relative to the additional cost. The corresponding routes are illustrated in Figure~\ref{fig:tartu_results}(c).

\section{Discussion and Conclusions}

This study explored how introducing a demand-weighted connectivity metric can influence the generation of public transport routes within AI-based optimization frameworks. Unlike previous works focused on improving algorithmic efficiency on synthetic benchmarks, our goal was to test whether adding connectivity-oriented equity measures changes the character or quality of generated networks.

The results show that the integration of $C_w$ slightly improves network balance and algebraic connectivity (robustness), but does not lead to dramatic performance gains. In practice, the main difficulty lies in preparing consistent and meaningful baseline data for real cities if such data is unavailable or simply does not exist — how to aggregate stops, estimate demand approximately, and convert spatial heterogeneity into graph form. Even with a strong algorithmic core, errors or simplifications in data preparation can easily distort the resulting routes.

Moreover, the networks obtained through optimization often appear irregular or unintuitive from an urban-planning point of view. Routes may zigzag, overlap, or fail to reflect real travel patterns, even if they score well on quantitative metrics. This points to a deeper problem: current evaluation methods for AI-generated transit networks remain narrowly metric-based, with little attention to spatial logic or human interpretability.

We therefore argue that progress in AI-assisted transit planning should focus not only on improving optimization objectives, but also on developing a transparent methodology for post-evaluating generated routes. Such a framework should combine quantitative indicators, spatial diagnostics, and expert assessment to determine whether a network is not only computationally optimal but also realistic, equitable, and usable for the city.

\vspace{.2em}

\section{Acknowledgments}

\bigskip
\noindent This work supported by the Ministry of Economic Development of the Russian Federation (IGK 000000C313925P4C0002), agreement No139-15-2025-010.

\bibliography{aaai2026}

\end{document}